\documentclass[useAMS,usenatbib,usedcolumn]{mn2e}

\def\HI{\ifmmode{\rm HI}\else{H\/{\sc i}}\fi}

\def\lsun{\ifmmode{{\mathrm L}_{\odot}}\else{L$_{\odot}$}\fi}

\def\msun{\ifmmode{{\mathrm M}_{\odot}}\else{M$_{\odot}$}\fi} 
\def\msunpc2{\ifmmode{{\mathrm M}_{\odot} \, {\mathrm{pc}}^{-2}}\else{M$_{\odot} \, {\mathrm {pc}}^{-2}$}\fi}

\def\kms{\ifmmode{{\mathrm{km \, s^{-1}}}}\else{${\mathrm{km \, s^{-1}}}$}\fi}

\def\aj{AJ}
\def\apj{ApJ}
\def\apjs{ApJS}
\def\aap{A\&A}
\def\mnras{MNRAS}
\def\pasp{PASP}

\usepackage[figuresright]{rotating}
\usepackage{lscape}

\usepackage{graphics}
\usepackage{epsfig}
\usepackage{multirow}

\usepackage{bigdelim}
\usepackage{bigstrut}

\newcommand{\sech}{\mathrm{sech} \,}

\title[The origins of S0s from PN kinematics]{Unravelling the origins of S0 galaxies using
maximum likelihood analysis of planetary nebulae kinematics}
\author[Cortesi et al.]
  {A.~Cortesi$^{1,2}$\thanks{email:ppxac2@nottingham.ac.uk}, M.~R.~Merrifield$^1$, 
    M.~Arnaboldi$^2$, O.~Gerhard$^3$, I.~Martinez-Valpuesta$^3$, 
  \newauthor  
   K.~Saha$^3$, L.~Coccato$^3$, S.~Bamford$^1$, N.~R.~Napolitano$^4$, P. Das$^3$, N.~G.~Douglas$^5$, 
  \newauthor  
   A.~J.~Romanowsky$^7$, K.~Kuijken$^6$,  M.~Capaccioli$^8$ and  K.~C.~Freeman$^9$ \\ 
   $^1$University of Nottingham, School of Physics and Astronomy, University
       Park, NG7 2RD Nottingham, UK \\
   $^2$European Southern Observatory, Karl-Schwarzschild-Strasse 2, 85748
       Garching, Germany \\ 
   $^3$Max-Planck-Institut f\"ur Extraterrestrische Physik,
       Giessenbachstrasse,  85741 Garching, Germany \\
   $^4$Istituto Nazionale di Astrofisica, Osservatorio Astronomico di
       Capodimonte, Via Moiariello 16, 80131 Naples, Italy \\
   $^5$Kapteyn Astronomical Institute, University of Groningen, PO Box 800,
       9700 AV Groningen, The Netherlands \\
   $^6$Leiden Observatory, Leiden University, PO Box 9513, 2300 RA Leiden, The
       Netherlands \\
   $^7$UCO/Lick Observatory, University of California, Santa Cruz, CA 95064, 
       USA \\
   $^8$Dipartimento di Fisica, Universit\`a ``Federico II'', Naples, Italy \\
   $^9$Research School of Astronomy and Astrophysics, Australian National
       University, Canberra, Australia} 

\begin{document}

\date{; }

\maketitle

\begin{abstract}
  To investigate the origins of S0 galaxies, we present a new method
  of analyzing their stellar kinematics from discrete tracers such as
  planetary nebulae.  This method involves binning the data in the
  radial direction so as to extract the most general possible
  non-parametric kinematic profiles, and using a maximum likelihood
  fit within each bin in order to make full use of the information in
  the discrete kinematic tracers.  Both disk and spheroid kinematic
  components are fitted, with a two-dimensional decomposition of
  imaging data used to attribute to each tracer a probability of
  membership in the separate components.  Likelihood clipping also
  allows us to identify objects whose properties are not consistent
  with the adopted model, rendering the technique robust against
  contaminants and able to identify additional kinematic features.

  The method is first tested on an N-body simulated galaxy to assess
  possible sources of systematic error associated with the structural
  and kinematic decomposition, which are found to be small. It is then
  applied to the S0 system NGC~1023, for which a planetary nebula catalogue has
  already been released and analyzed by \citet{EDO}.  The correct
  inclusion of the spheroidal component allows us to show that,
  contrary to previous claims, the stellar kinematics of this galaxy
  are indistinguishable from those of a normal spiral galaxy,
  indicating that it may have evolved directly from such a system via
  gas stripping or secular evolution.  The method also successfully
  identifies a population of outliers whose kinematics are different
  from those of the main galaxy; these objects can be identified with a
  stellar stream associated with the companion galaxy NGC~1023A.
\end{abstract}


\section{Introduction}
\label{sec:introduction}

Lenticular or S0 galaxies lie between ellipticals and spirals on the
Hubble sequence, since they have the featureless old stellar
populations of elliptical systems, but also contain the disk
components associated with spirals.  They thus potentially represent a
key element in attempts to understand the relationship between the
main types of galaxy, but at the moment there is no consensus as to
the end of the Hubble sequence to which they are most closely related.
Clearly, some process has consumed their gas and shut off their star
formation.  However, this termination could be the result of galaxy
mergers, much in the manner that star formation is believed to be
quenched in elliptical galaxies, or it could arise from some much
gentler process, such as ram pressure stripping, which simply removes
the gas from normal spiral galaxies.

Perhaps the best way of discriminating between these scenarios is
offered by the stellar dynamics of S0 galaxy disks.  If they are the
product of relatively gentle gas-stripping processes, one would expect
the stellar dynamics to be unaffected, and should be identical to
those of the progenitor spiral system, with rotation dominating
relatively modest amounts of random motion in the disk
\citep{Salamanca, Williams}.  If, however, the S0s are the product of
minor mergers, with a mass ratio higher than 10:1, one would expect
the merging process to be imprinted in the stellar dynamics, with
random velocities as high or even higher than the rotational
velocities at all radii \citep{bour2}.

One difficulty in using such a diagnostic is that it requires accurate
stellar kinematics measured in the outer parts of the galaxy where the
disk dominates, and at low surface brightness, making conventional
absorption-line spectroscopy difficult.  However, planetary nebulae
(PNe) provide an excellent probe of stellar kinematics in this regime:
they are easily detected and can have their velocities measured from
their characteristic strong emission lines, and they have proved to be
an unbiased tracer of the global stellar population \citep{Ciardullo1,Ciardullo2,Napolitano1,Coccato}.  We have designed and built a special-purpose
instrument, the Planetary Nebula Spectrograph [PN.S; \citet{Nigel}],
specifically to exploit this tracer of stellar kinematics at low
surface brightnesses.  Although originally intended to study
elliptical galaxies \citep{Romanowsky,Coccato,Napolitano2}, PN.S has already proved very
effective in exploring the disks of spiral galaxies \citep{Merrett},
so an obvious next step was to use it to probe the kinematics of S0
systems.

As a pilot study, we observed the relatively nearby S0 galaxy NGC~1023
with PN.S.  The observations and the resulting catalogue of positions
and line-of-sight velocities of PNe are described in \citet{EDO}.  The
analysis in that paper concluded that NGC~1023 has very peculiar
kinematics, giving the appearance of a normal rotationally-supported
disk galaxy at small radii, but having entirely random velocities at
large radii, inconsistent with either of the expected scenarios.
However, the relatively simple dynamical analysis that led to these
conclusions had some significant shortcomings.  First, it assumed that
the light at large radii was dominated by the disk component, and
neglected any contribution from the spheroid.  As we will see, this
assumption can lead to sizable systematic errors in the inferred disk
kinematics.  Second, it calculated dynamical quantities such as
velocity dispersions by binning the data both azimuthally and radially
in the galaxy.  Although such binning provides an adequate signal from
a sparsely-sampled velocity field and a non-parametric measure of the
local kinematics, as we will show below the kinematic properties vary
continuously and quite rapidly with azimuth, so such binning averages
away significant amount of information that is present in the raw
data.  The binning process also makes it difficult to identify any
contamination from PNe that are either unrelated background objects,
or lie within the system but do not match the expected kinematics.

In this paper, we therefore present a somewhat more sophisticated
analysis designed to circumvent these shortcomings.  While still
binning the data in the radial direction, so as to extract a general
non-parametric view of the galaxy's dynamics, we apply a maximum
likelihood analysis within each radial bin, so as to extract the
maximum amount of information from the azimuthal variations in
kinematic properties.  We also model both disk and spheroid in the
system, and use a full two-dimensional photometric decomposition of
the galaxy, which allows us to allocate to each PN a probability as to
the component to which it belongs.  

The method we have developed to construct such a model is described in
Section~\ref{sec: The method}, and in Section~\ref{sec: Application to
  a model galaxy} we test the technique on a simulated galaxy for
which we have prior knowledge of the structural components and
kinematics. In Section~\ref{sec: NGC1023 application} we make a first
real application of the method to the NGC~1023 catalogue, in which we
show that this likelihood fitting can reproduce the strange results in
\citet{EDO} if we also neglect the spheroidal component, but that the
more complete model presented here results in a picture in which
NGC~1023 has a normal rotationally-supported disk.  We also illustrate
the power of likelihood fitting by showing how outlier points can be
identified in the catalogue, and associated with a minor on-going
merger.  The conclusions are presented in Section~\ref{sec:
  Conclusions}.

\section{Kinematic Likelihood Fitting}
\label{sec: The method}

In this section we give a brief introduction to likelihood analysis,
before going on to the specifics of fitting particular galaxy models.
In Section~\ref{sec:disk}, we present the case of a pure disk
galaxy, as it provides both a simple example of the technique and a
useful point of comparison to previous analyses that have assumed that
the disk is the dominant component.  In
Section~\ref{sec:diskandspheroid}, we proceed to the more realistic
situation in which we model both disk and spheroidal component.  In
Section~\ref{sec:decompose} we discuss how the individual discrete
kinematic tracers can be assigned, at least in a probabilistic sense,
to either the disk or the spheroid using photometric data.  

In general, given a set of $N$ independent velocity measurements,
$v_{i}=(v_{1},...,v_{N})$, drawn from a probability density function
$F(v_{i};\vec{\theta})$, where $\vec{\theta}=(V,\sigma)$ is a
set of parameters whose value is unknown, the likelihood function is given by
\begin{equation}
\mathcal{L}= \prod_{i} F(v_{i};\vec{\theta}).
\label{eq:likelihood}
\end{equation}
The values of ${V}$ and ${\sigma}$ that maximize $\mathcal{L}$
are the best estimators for the true values of these parameters.
Moreover the method of maximum likelihood coincides with the method of
the least squares in the special case of a set of $N$ Gaussian
distributed independent measurements, in which case the likelihood
function is directly related to the usual $\chi^2$ statistic,
\begin{equation}
\Delta\chi^{2}(\vec{\theta})=-2 \Delta \ln \mathcal{L}(\vec{\theta}).
\end{equation}
Thus, in this case it is straightforward to determine the confidence
region around the best estimator:
\begin{equation}
\ln \mathcal{L}(\vec{\theta}) \ge \ln \mathcal{L}_{max} - \Delta \ln \mathcal{L},
\end{equation}
where values of $\Delta \chi^{2} $ or $-2 \Delta \ln \mathcal{L}$,
corresponding to a desired confidence limit, for joint estimation of
$m$ parameters, are readily available in tabulated form. In this paper
we use a $\Delta \chi^{2}$ that corresponds to a $1 \sigma$ coverage
probability.  In particular, $\Delta \chi^{2}=5.39$ for $m=4$, and
$\Delta \chi^{2}=4.11$, for $m=3$.

Finally, once the best-fit parameters have been found, one can
calculate the contribution of each data point to the likelihood given
this optimum model.  To quantify this contribution in terms of whether
the kinematic tracer in question is consistent with being drawn from
the best-fit model, one can generate a large number of individual
velocities drawn from a velocity distribution matching the model, and
see at what percentile of the distribution the data point lies.  We
can thus identify any objects whose velocities lie outside this
confidence interval, which are most likely interlopers.  By rejecting
these data points and iterating on the fit, we can render this process
robust against a small amount of contamination, in a straightforward
generalization of the $3\sigma$ clipping procedure \citep{Merrett2003}
used in determining membership of velocity distributions that are
assumed to be Gaussian. We define the {\it likelihood clipping
  probability threshold} as the value at which we cut the
distribution, and we reject all the PNe beyond this limit.

In principle, such likelihood fitting could be applied to a full
dynamical model across an entire galaxy, at all radii and azimuths.
However, in this analysis we are interested in leaving as much freedom
as possible in the resulting radial profile of dynamical properties,
so as to explore the full range of possibilities predicted by
different models of S0 formation.  We therefore adopt a hybrid
approach in which the data are binned into elliptical annuli
matched to the inclination of the disk, to extract points over a range
of intrinsic radii.  This binning allows each section of the disk to be
modeled as an independent non-parametric data point.  However in the
azimuthal direction around each bin we use a full likelihood analysis
that accurately models the expected variation in the line-of-sight
velocity distribution with azimuth, as derived from very general
dynamical considerations.

\subsection{Likelihood analysis for a disk model}\label{sec:disk}

Consider a general rotating disk model for a galaxy.  Part of the
line-of-sight velocity of each object within it, $v_{i}$, is the
projection of the galaxy's mean rotation velocity $V$, at an azimuthal
angle $\phi$ within the galaxy, corrected for the inclination of the
galaxy to the line of sight $i$, and for the systemic velocity of the
galaxy itself $V_{sys}$:
\begin{equation}
V_{los}=V_{sys}+V \sin(i) \cos(\phi).
\end{equation}

Superimposed on the net rotational velocity are the random motions of
the individual stars, which can be quantified by their velocity
dispersion in different directions.  For an axisymmetric disk, these
components are most naturally expressed in cylindrical polar
coordinates aligned with the axis of symmetry, ($R,\phi,z$), and the
line-of-sight velocity dispersion, $\sigma_{los}$, is made up from a
projection of these components such that
\begin{equation}
\sigma^{2}_{los}=\sigma^{2}_{r} \sin^2 i \sin^2\phi +\sigma^{2}_{\phi} \sin^2i \cos^2\phi + \sigma^{2}_{z} \cos^2i
\end{equation}
Generally speaking, in disk galaxies,  $\sigma_z$ is the smallest component of the
velocity dispersion, and in a nearly edge-on galaxy, like the ones we are focusing on, it  does not
project much into the line-of-sight velocity dispersion, so the value
of  $\sigma_{los}$ is dominated by the other two components. As
can be seen from the above equation, $\sigma_{los}^2$  varies sinusoidally such that its
value is set by $\sigma_R$ on the minor axis and $\sigma_\phi$ on the
major axis.  Thus, by fitting the variation in $\sigma_{los}$ with
azimuth one can determine the values of both of these components of
the galaxy's intrinsic velocity dispersion.

If we adopt the simplest possible model in which the line-of-sight
velocity distribution is Gaussian at every point with a mean velocity
of $V_{los}$ and a dispersion of $\sigma_{los}$, we now have the
requisite form for the probability density function, 
\begin{equation}
F(v_{i};V,\sigma_{r},\sigma_{\phi}) \propto \exp \left[ - \frac{[v_{i} - {V_{los}(V)}]^{2}}{2 [\sigma_{los}(\sigma_r,\sigma_\phi)]^{2}} \right].
\end{equation}
The values $V, \sigma_{r}, \sigma_{\phi}$ that maximize $\mathcal{L}$
are the best estimators for the true values of these kinematics
parameters within each bin.

\subsection{Likelihood analysis for a disk $+$ spheroid
  model}\label{sec:diskandspheroid}

In addition to the disk component, most systems also usually contain a
spheroidal component comprising either a central bulge or an extended
halo (or both).  Indeed, one of the traditional defining features of
S0 galaxies is that they often have rather prominent bulges.  A
kinematic model without such a component may therefore be
significantly incomplete; worse, the amount of bulge light varies with
azimuth around the galaxy, since close to the minor axis in a
highly-inclined system the bulge is usually the dominant component, so
one might expect the variation in velocity dispersion with azimuth,
used above to extract the different components of disk velocity
dispersion, to be systematically distorted.

One slight complication in adding in such a spheroidal component is
that it does not have the same symmetry properties as the disk
component.  Thus, the elliptical annuli that contain data from a small
range in radii in the disk does not correspond to the same range in
radii in the more spherical spheroidal component, but samples a larger
range of radii both due to the difference in shape and the effects of
integration along the line of sight.  However, the variation in
kinematics with radius in such a hot component is generally rather
slow, so the greater averaging in radius imposed by the choice of
elliptical annuli should not have a major impact.  Further, the
validity of this assumption of slow variation with radius can be
tested a posteriori by seeing how the inferred kinematics of the
spheroid change from bin to bin.

Accordingly, we adopt the simplest possible model for the kinematics
of the spheroidal component, in which the line-of-sight velocity
distribution is assumed to be a Gaussian with zero mean velocity
(relative to the galaxy). We have thus assumed that any rotation in
the spheroidal component is negligible, although, as we shall see
below, this assumption can also be relaxed.  The velocity dispersion,
$\sigma_{s}$, is left as a free parameter to be modeled by the data.
The only other parameter that we need to specify is the probability
that each individual tracer object at its observed location belongs to
the spheroidal component, $f_{i}$, so that the full probability
density function can be written
\begin{eqnarray}
\lefteqn{F(v_{i};V,\sigma_{r},\sigma_{\phi},\sigma_{s}) \propto  
      {f_{i} \over \sigma_s} \exp \left[ - \frac{v_{i}^{2}}{2 \sigma_{s}^{2}} \right]}  \nonumber \\ 
 & & + {{1-f_{i}} \over \sigma_{los}}\exp \left[ - \frac{(v_{i} - {V_{los}})^{2}}{2\sigma_{los}^{2}} \right], 
\label{eq:dplussmodel}
\end{eqnarray}
where the velocity dispersions in the denominators of the term in
front of each Gaussian ensure that this function is properly normalized.
From this velocity distribution one can then construct a likelihood function for a
particular set of model parameters using Equation~\ref{eq:likelihood}.
The values of $V, {\sigma}_{R}, {\sigma}_{\phi},
{\sigma}_{s}$ that maximize $\mathcal{L}$ then provide the
estimators for these parameters in the best-fit model.

\subsection{Spheroid--disk decomposition}\label{sec:decompose}
The only parameters that we have not yet specified are the $f_i$
values, which define the probability that a given object is in the
spheroid.  We cannot solve for these quantities directly in the
likelihood maximization since, as noted above, the relative
contributions of spheroid and disk vary with azimuth within a single
bin, so each kinematic data point has its own individual value.
Fortunately, we have additional information that has not yet been
used.  In particular, we can apply a two-dimensional galaxy fitting
routine such as GALFIT \citep{Peng} to imaging data in order to
decompose the starlight into spheroid and disk components.  The fit to
the light distribution then provides a direct estimate of the fraction
of the starlight from each component at the location of any stellar
tracer.

This approach for determining the decomposition into bulge and disk
components has the great advantage that the broad-band imaging data
offer a much less sparse sampling of the stellar spatial distribution
than the PNe, providing an intrinsically more accurate answer.
Furthermore, there are significant selection effects in detecting PNe, as
they are harder to find against the bright background of the inner
parts of a galaxy, so their spatial arrangement is not an unbiased
representation of the stellar distribution.  Note, however, that this
bias is a purely spatial one, in that all line-of-sight velocities are
equally detectable, so their use in the kinematic analysis is not in
any way compromised.

The only slight subtlety in applying such analysis to PNe is that
their number per unit stellar luminosity has been shown to vary
systematically with the colour of the population \citep{Ciardullo3},
with a lower PN density per unit galaxy luminosity for redder objects
\citep{Buzzoni}.  If there is a difference in colour between the disk
and spheroidal components, then one has to apply a correction
in order to convert the fraction of spheroidal light at any given
point into the probability that a PN detected at that point belongs to
the spheroid.  In practice, such colour terms can be straightforwardly
determined by performing the decomposition on images taken in
different bands, and using the resulting colours of the different
components to correct the probability.

\section{Application to a simulated galaxy}
\label{sec: Application to a model galaxy}

As a first test of this method, we perform the fit to a model galaxy
obtained from a self-consistent N-body simulation. The simulation was
provided by Kanak Saha (private communication). This simulated galaxy
is constructed using a nearly self-consistent bulge-disk-dark halo
model \citep{Kuijken, Widrow} to mimic a typical lenticular
galaxy. The value of Toomre $Q$ is high enough to prevent strong
$2$-armed spirals from forming in the disk. The spheroid follows a
Sersic profile with an index of $3.5$, while the disk is exponential
with a vertical $\sech^{2}$ profile. The dominant dark matter halo
specifies the system's rotation curve. We chose to ``observe'' this
galaxy at an inclination of $60$ degrees. For the following tests, we
set the likelihood clipping probability threshold to $2.1 \sigma$.

\subsection{Testing the likelihood method using  {\it a priori} knowledge of PN positions}

\begin{figure}
\includegraphics[width=0.45\textwidth]{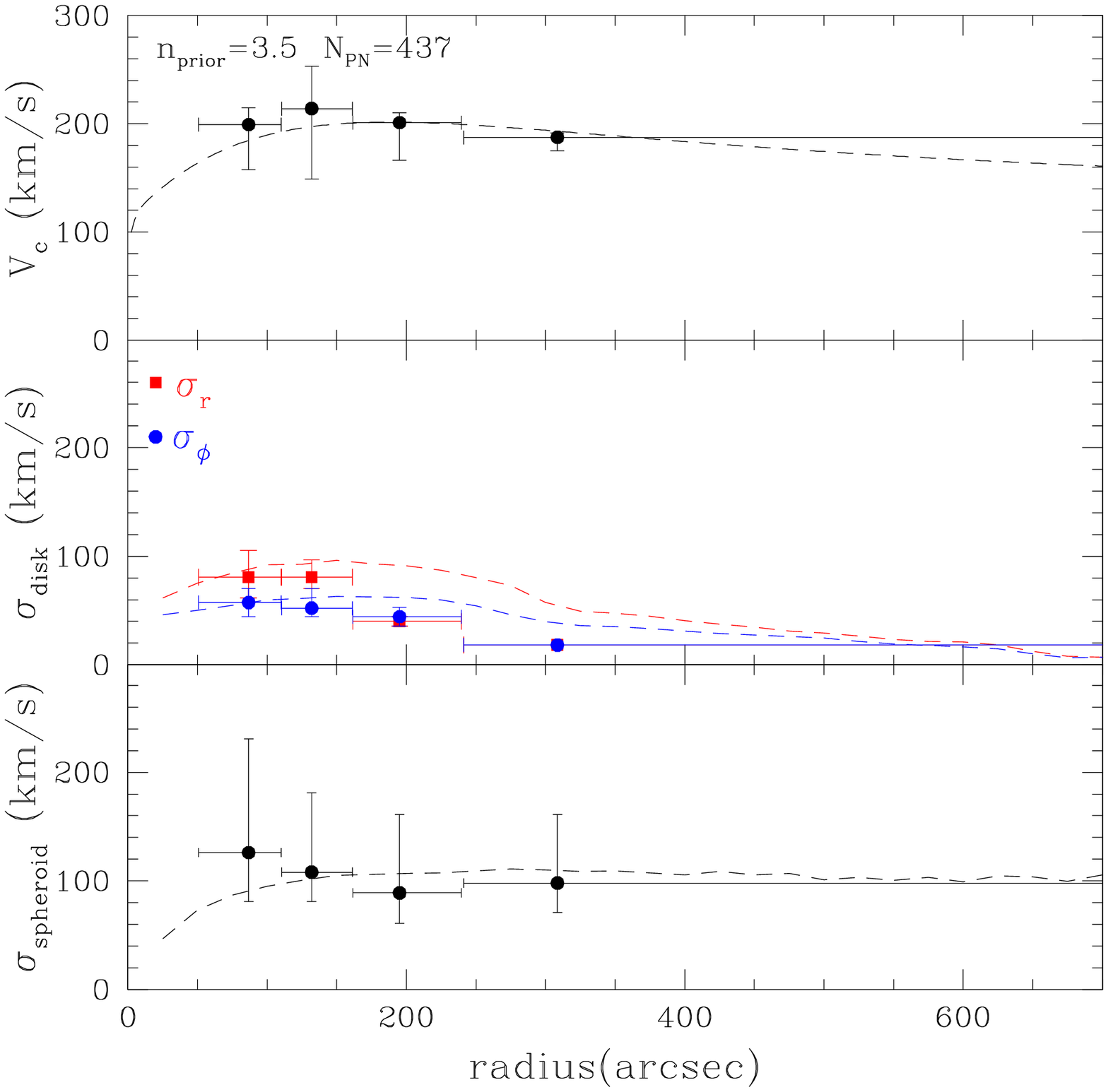}
\includegraphics[width=0.45\textwidth]{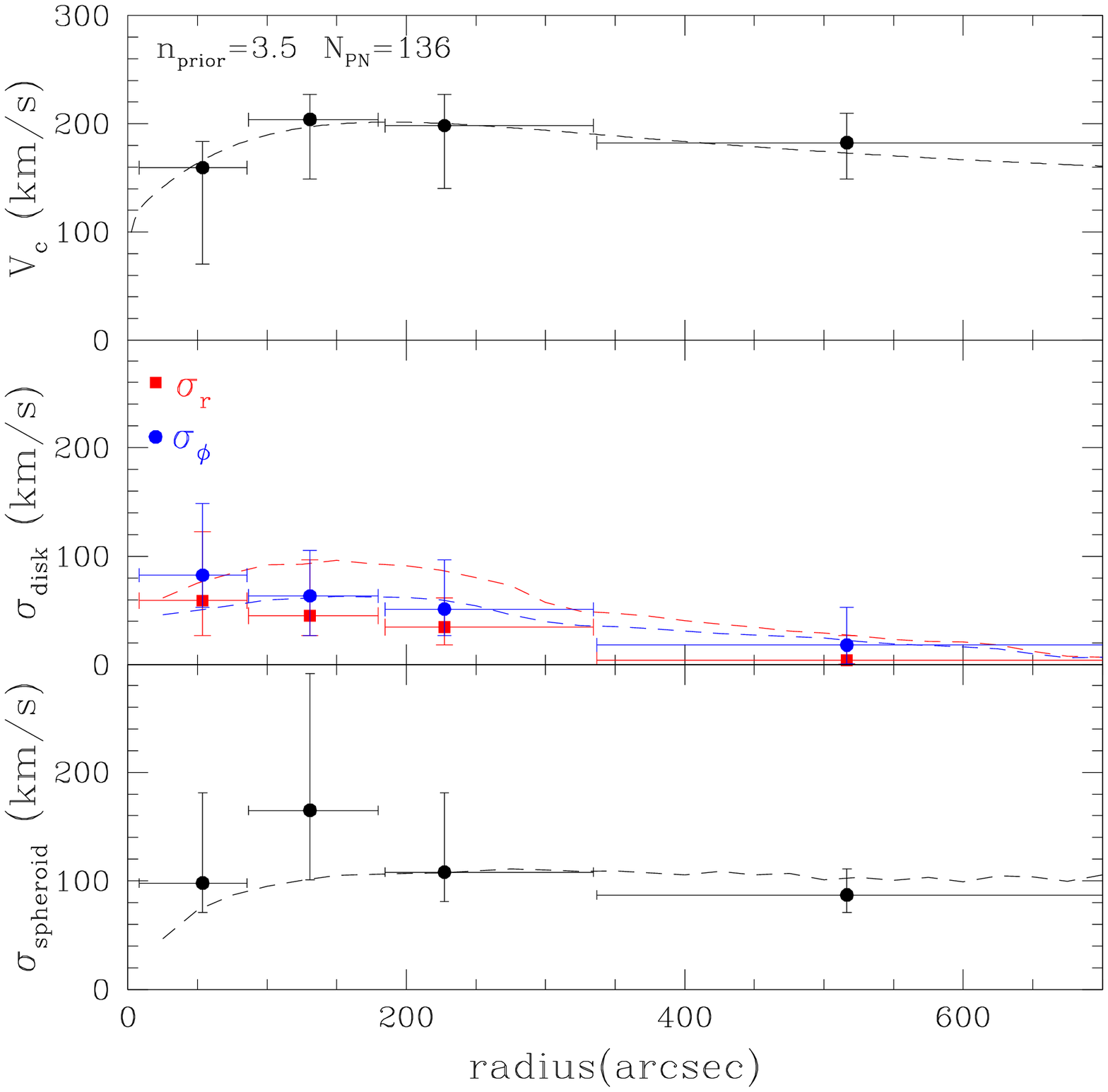}
\caption{Derived circular velocity and the components of the
  velocity dispersion versus radius for the model galaxy in the disk and in the
  bulge, for $437$ particles, upper panel, and $136$ particles, bottom panel.The filled symbols
  are from the maximum-likelihood analysis, with vertical error bars
  indicating uncertainty and horizontal error bars showing the extent
  of each radial bin (with the point plotted at the median radius
  for a PN in that bin). The dashed  lines represent the model circular  velocity, velocity dispersion in the disk and in the spheroid.\label{fig:model1}}
\end{figure}

In a simulated galaxy, we have access to inside information which
allows us to test specific aspects of this modeling process.  In
particular, the ``stars'' are tagged according to whether they are
members of the disk or spheroid, so we can assign the objects to
specific components with certainty, rather than using the
probabilistic approach described above.  Accordingly, we have carried
out the likelihood analysis without the probabilistic decomposition,
with the results presented in Figure~\ref{fig:model1}.  The upper
panel shows the results obtained with a generous catalogue of 437
objects, while the lower panel shows how well we can do with only 136
kinematic tracers.  Although not strictly valid for kinematically-hot
S0 galaxies, an asymmetric drift correction has been applied to
convert the derived rotation velocity into a circular velocity for
direct comparison with the simulation's known functional form, using
the formula
\begin{equation}
V_{c}^{2} = V^{2} + \sigma_{\phi}^{2} - \sigma_{r}^{2} \left(1 - {r_{h} \over
r_{D}} + \frac{d \ln \sigma_{r}^2}{d \ln r}\right),  
\end{equation} 
where $V_{c}$ is the circular velocity, $r_{h}$ is the median radius
in each bin, and $r_{D}$ is the disk scale length \citep{Kormendy}. The final gradient term of the equation has been estimated assuming that $\sigma_r ^{2}$ follows an exponential profile and performing a linear fit between $\ln \sigma_r ^{2}$ and the radius.  

While the errors are, as expected, larger for the smaller sample,
there are no systematic differences between them, and both do a good
job of recovering the simulation's circular velocity. It is
interesting to see that $\sigma_{R}$ is moderately but systematically
too low for the small sample-size, and remains low at large radii even
for the larger sample.  Since the main remaining assumption in this
model is that the velocity distributions of the individual components
are intrinsically Gaussian, it seems likely that this modest
systematic error occurs due to the breakdown of this assumption.
However, we note that $\sigma_{\phi}$ and $v$ are both quite
accurately recovered, so we can estimate with some confidence the
balance between random and rotational motion, $\sigma_\phi/v$, which
is the main physical quantity that we are seeking to use as a
diagnostic in this analysis of the origins of S0 galaxies.

\subsection{Testing the effect of the photometric spheroid--disk decomposition}

\begin{figure}
\includegraphics[width=0.45\textwidth]{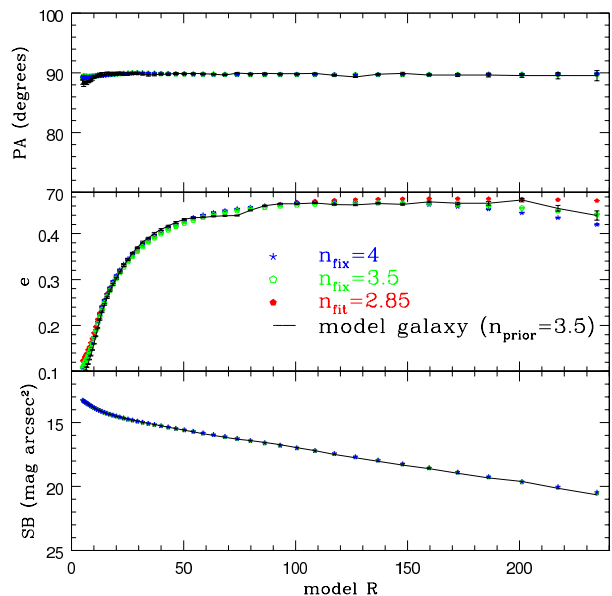}
\caption{Results from the isophotal analysis of the simulated galaxy, black filled triangles, and the galaxy models obtained with GALFIT leaving the Sersic index $n$ as a free parameter, red filled pentagons, imposing $S_{GALFIT}=3.5$, green open pentagons and imposing $S_{GALFIT}=4$, blue starred symbols. The top row shows the fitted position angles, the middle row the ellipticity and the bottom row the surface brightness. \label{fig:comp}}
\end{figure}

We now introduce the remaining aspect of the full model-fitting
process, the spheroid--disk decomposition, which allows us to assign
a probabilistic membership of each kinematic tracer to each component,
since, as discussed above, in a real galaxy we do not have the luxury
of this knowledge {\it a priori}.  We constructed an
image of the simulated galaxy and convolved it with a suitable Gaussian
point-spread function. We then used GALFIT to model the resulting
simulated broad-band image.  The principal parameters returned by this
fitting process are the spheroid and disk magnitudes, their respective
scale lengths, the Sersic index of the spheroid, and the flattening
and position angle of the components.

Interestingly, if the parameters are all left free, then the spheroid
is found to have a best-fit Sersic index of 2.8, significantly smaller
than the known value for this simulation of 3.5.  The value of the
effective radius is also found to be systematically too small.
However, an almost equally-good fit is found if we fix the Sersic
index to 3.5.  In fact, this fit is by some measures superior:
Figure~\ref{fig:comp} shows the values of position angle, ellipticity
and surface brightness obtained by fitting elliptical isophotes to both the
simulated galaxy image and the models in which the Sersic index is
either fixed or left free.  While both models reproduce the 
position angle and surface brightness  equally well, the ellipticity is clearly better fitted by the model
with the Sersic index fixed at the right value.  This conflicting
information underlines the complexity of non-linear model fitting, and
illustrates its basic limitations.  

\begin{figure*}
\begin{tabular}{lr}
\includegraphics[width=0.45\textwidth]{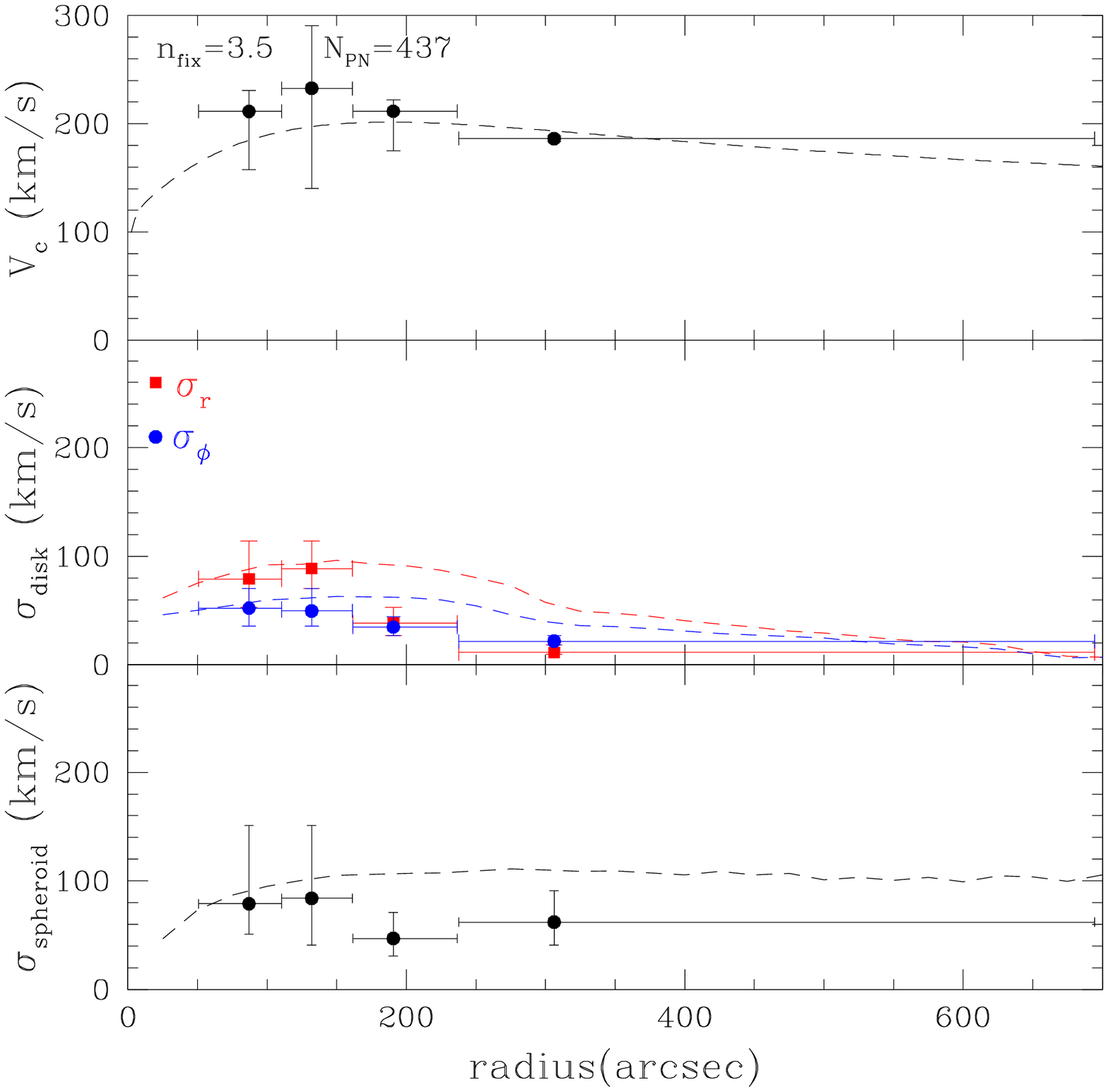}
\includegraphics[width=0.45\textwidth]{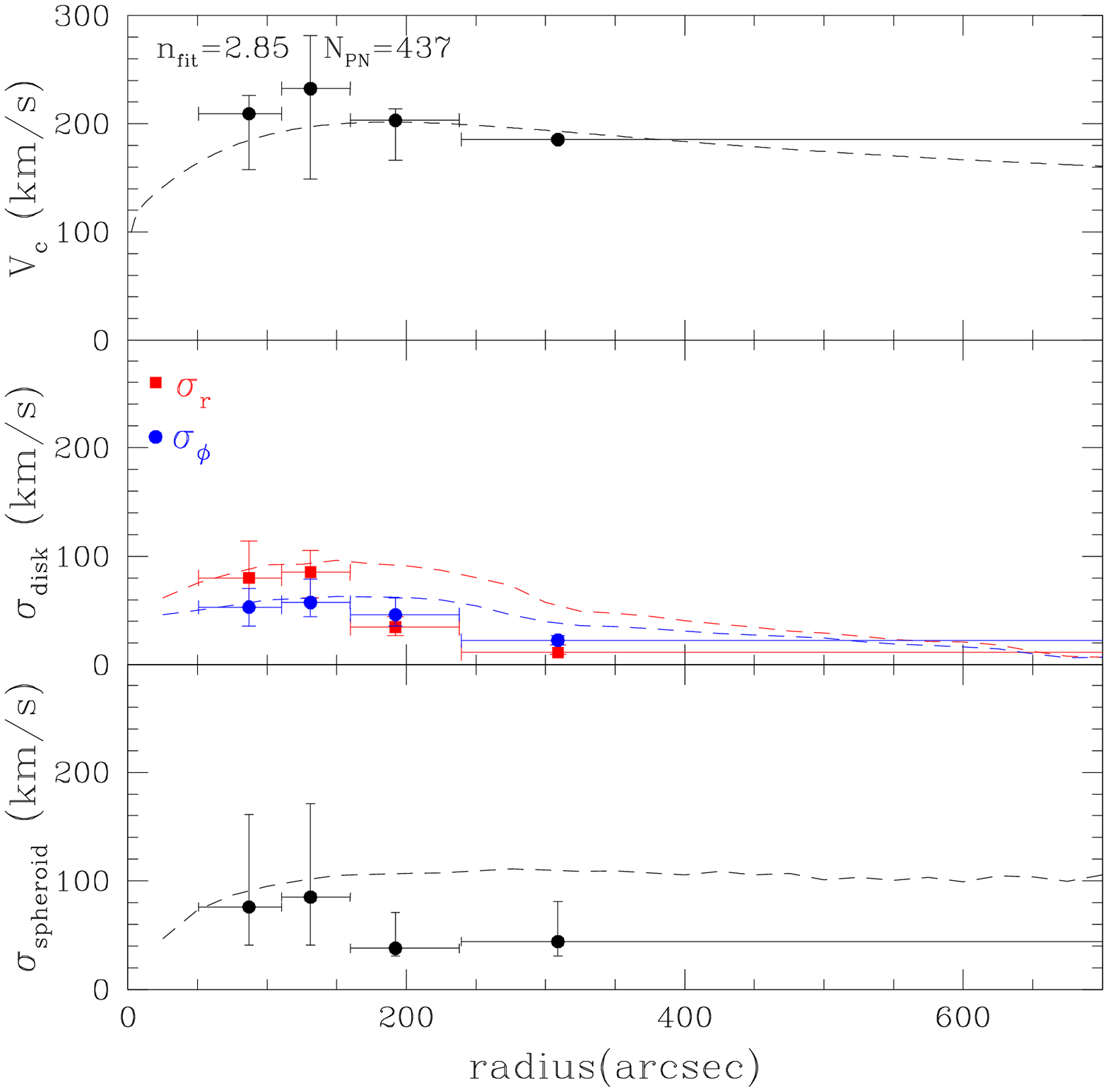}\\                   
\includegraphics[width=0.45\textwidth]{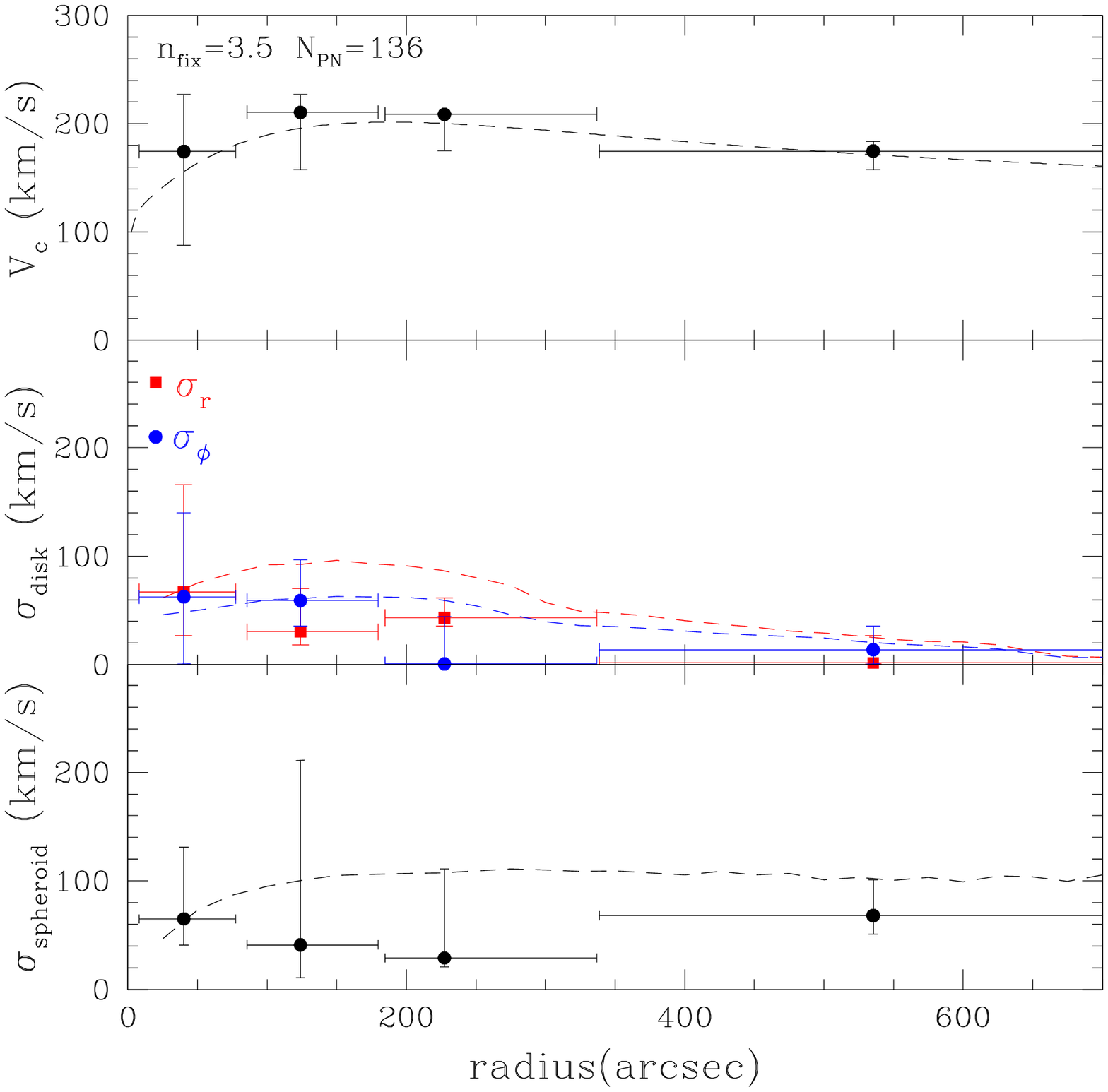}
\includegraphics[width=0.45\textwidth]{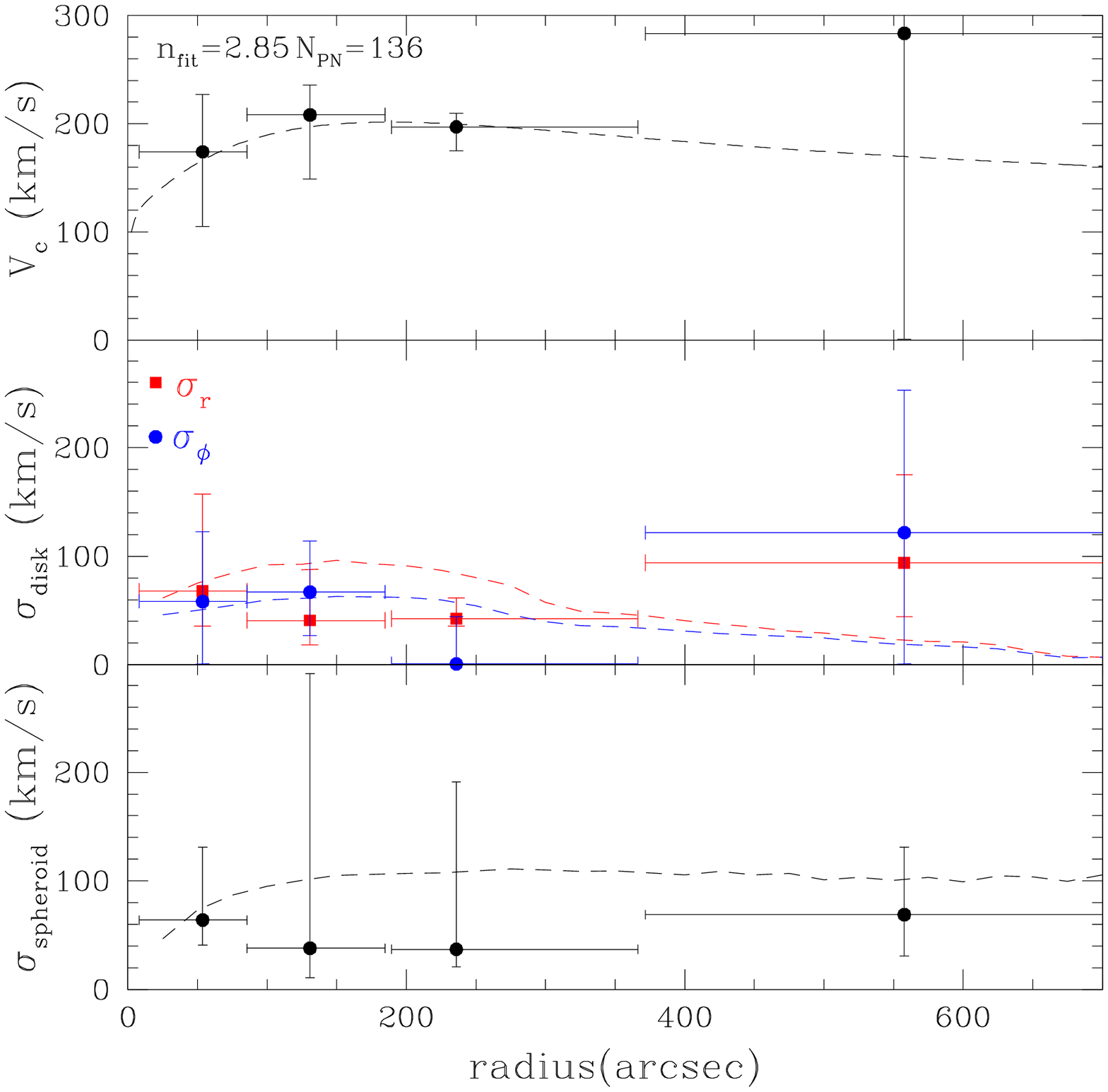}
\end{tabular}
\caption{Derived circular velocity and the components of the velocity
  dispersion versus radius for the model galaxy in the disk and in the
  bulge.  The left panels show the result when the spheroid's Sersic
  index is fixed at its true value of 3.5, while the right panels show
  the results when it is left as a free parameter.  Upper panels are
  for a larger catalogue of 437 kinematic tracers; lower panels for a
  smaller sample of 136.  The filled symbols are from the
  maximum-likelihood analysis, with vertical error bars indicating
  uncertainty and horizontal error bars showing the extent of each
  radial bin (with the point plotted at the median location for a
  kinematic tracer in that bin). \label{fig:model3}}
\end{figure*}

Of course, for a real galaxy, we would not know the ``right'' value
for the Sersic index, so would not be able to choose between these
models.  Such systematic errors in GALFIT fitting therefore pose a
potential limitation to the effectiveness of the modeling procedure
set out in this paper, if the resulting kinematics turn out to depend
sensitively on the spheroid--disk decomposition.  To assess the
impact of such effects, we carried out the maximum likelihood modeling
using both of these decompositions.  We also checked the effect of
kinematic tracer sample size by simulating catalogues of 437 and 136
PNe.  The resulting kinematics are presented in
Figure~\ref{fig:model3}.  The good news is that the results are
largely insensitive to systematic errors arising from the disk --
spheroid decomposition.  For the smaller catalogue, the extra
uncertainty that arises from the decomposition process mean that the
errors in the derived kinematic quantities become quite large, but
there is no evidence for any systematic error in the process.  Thus,
it appears that this maximum likelihood procedure is quite reliable,
and robust against the most likely sources of systematic error in the
spheroid--bulge decomposition.

\section{Application to NGC~1023}
\label{sec: NGC1023 application}

\begin{figure}
\includegraphics[width=0.45\textwidth]{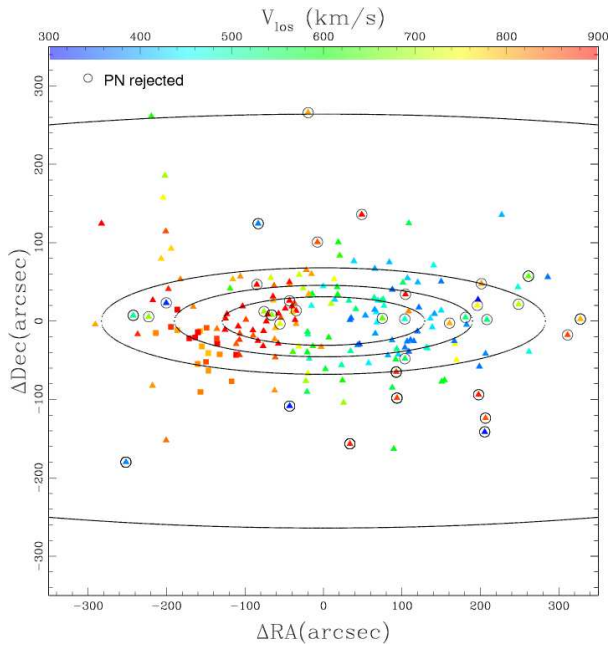}
\caption{Positions and velocities of the PNe detected in NGC~1023.
  Colors bar indicates velocities.  The elliptical annuli show the radial
  binning of the data.  The circled points are those rejected in the
  pure disk model,  with the likelihood clipping probability threshold increased to $1.65 \sigma$.  Squares show PNe associated with the companion
  galaxy NGC~1023A \citep{EDO}.\label{fig:2gal-nh}}
\end{figure}

Having developed this methodology for extracting kinematic properties
of disk galaxies from individual stellar tracers, and tested its
reliability, we can now apply it to the existing PNe data for the S0
galaxy NGC~1023.  The catalogue of 183 PNe positions and velocities,
published by \citet{EDO}, are presented in Figure~\ref{fig:2gal-nh}.
Note the ``hole'' in the middle of the galaxy resulting from the
difficulty of detecting PNe against the bright stellar continuum at
these radii, as discussed above.  Application of
the new method to these data has the benefit that they have already
been studied using a more conventional tilted-ring model \citep{EDO},
with which our results can be compared, plus the somewhat peculiar
kinematics apparently found in this system warrant further
investigation.  Since the previous analysis by \citet{EDO} neglected
any contribution to the kinematics from the bulge, we begin by
considering the disk-only model to make a direct comparison.

\subsection{The disk model}

\begin{figure}
\includegraphics[width=0.45\textwidth]{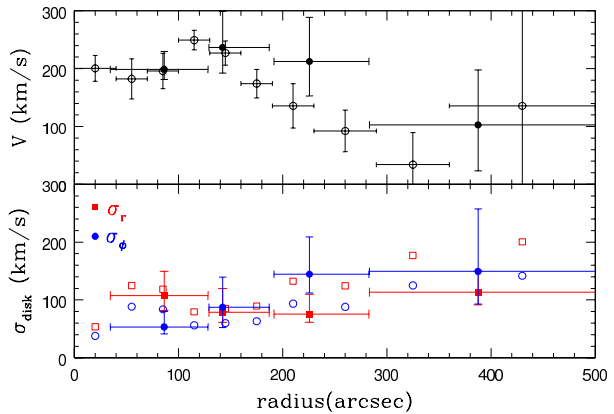}
\caption{Derived mean rotation speed and the components of the
  velocity dispersion versus radius for NGC~1023, for the case of a
  disk-only model. The filled symbols are from the maximum-likelihood
  analysis, with vertical error bars indicating uncertainty and
  horizontal error bars showing the extent of each radial bin (with
  the point plotted at the median location for a PN in that bin).  The
  open symbols reproduce the results of the tilted-ring analysis by
  \citet{EDO}.\label{fig:like-nh}}
\end{figure}

The application of the above approach, in which the data are binned
radially (into the annuli indicated in Figure~\ref{fig:2gal-nh}) but
modeled azimuthally by likelihood analysis, is presented in
Figure~\ref{fig:like-nh}.  For this analysis and the following
sections, we have estimated the galaxy's inclination from the
ellipticity of the disk component derived from the two-dimensional fit
to the photometry (see Section~\ref{sec:phot}); assuming that the
disk is intrinsically axisymmetric, we infer a value of $i=74.3$
degrees.  Likelihood clipping has been applied such that PNe with a
probability of less than $2.1 \sigma$ of being drawn from the disk
model are rejected.  As is clear from this figure, the analysis
largely reproduces the peculiar result of \citet{EDO}, with an
inferred rotation speed that falls rapidly outside 300~arcseconds,
accompanied by a rise in velocity dispersion.  As previously noted,
such kinematics were not predicted by any of the simpler scenarios for
S0 formation, therefore motivating this further investigation.

One immediate clue is offered by the only significant difference
between the results of the two analyses presented in
Figure~\ref{fig:like-nh}.  Specifically, while the radial dispersion
exceeds the tangential component in the \citet{EDO} analysis, the
opposite is the case in the maximum likelihood fit.  In fact, the
ordering of dispersions was fixed in the first analysis, as their
ratio was set by the epicyclic approximation, which forces $\sigma_{r}
> \sigma_{\phi}$ unless the rotation curve is very rapidly rising
\citep{Binney}.  In the current analysis, we have left both
components as free parameters, and it is telling us that we then find
$\sigma_{\phi} > \sigma_{r}$, which is not what the
physically-motivated epicyclic approximation would produce, suggesting
that there is some basic flaw in the model.

As noted above, the principal missing element in this model is the
spheroidal component, which is consistent with the enhanced value of
$\sigma_\phi$.  Specifically, a spheroid will contribute a relatively
small number of PNe with a large dispersion but zero mean velocity.
Close to the minor axis of the galaxy, the disk PNe will also have
zero mean line-of-sight velocity, whereas on the major axis their
distribution will be offset in velocity due to rotation.  Combining
the two distributions with different mean velocities will result in a
larger incorrectly-inferred dispersion than combining them where
their mean velocities are the same.  Since the tangential component
projects mostly into the line-of-sight close to the major axis, its
derived value would be more enhanced by this contamination than that of the
radial component, consistent with the results in
Figure~\ref{fig:like-nh}.  

\begin{figure}
\includegraphics[width=0.45\textwidth]{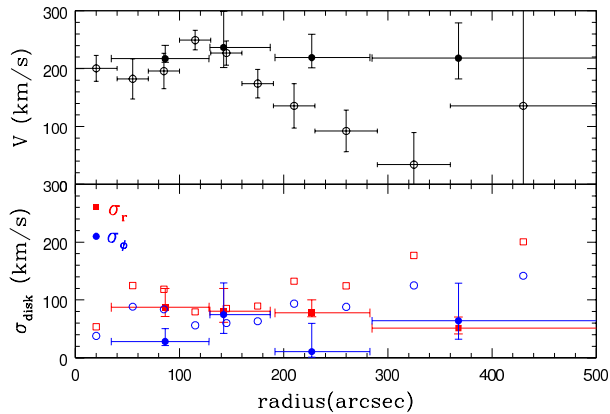}
\caption{As for Figure~\ref{fig:like-nh}, but with the likelihood
  clipping probability threshold increased to $1.65 \sigma$.\label{fig:like-nh-clip}}
\end{figure}
A very simple test of whether such contamination could be responsible
for the unphysical results can be made by increasing the severity of
the likelihood clipping, to try to remove the contaminants from the
fit.  Figure~\ref{fig:like-nh-clip} shows the success of such a
process, in which a more aggressive threshold resulted in 34 PNe being
rejected.  The components of the velocity dispersions are now ordered
in the manner physically expected for a disk population.  Further, the
bizarre behavior of the kinematics has entirely vanished, with the
rotation velocity now remaining approximately constant out to large
radii, and the dispersion remaining low, just as one would expect
for a normal disk population.

As further evidence that the cause of the contamination is the
spheroid, Figure~\ref{fig:2gal-nh} highlights the locations of the PNe
rejected in this iterative clipping.  The PNe appear spread throughout
the galaxy, and not flattened into the aspect ratio of the disk,
indicating that they are probably not the result of a poor disk model.
There is, however, some indication of an asymmetry in rejected PNe
between the two sides of the galaxy, which we return to in
Section~\ref{sec: Rejections}.  

Clearly we need to model a spheroidal component as well as the disk,
but it is also heartening to see that the likelihood rejection method
does a respectable job of dealing with even such a high level of
contamination, in which almost $20\%$ of the PNe do not seem to be
drawn from the assumed model.  This result provides some confidence in
the robustness of the adopted procedure.

\subsection{Spheroid -- disk decomposition}\label{sec:phot}

 \begin{figure*}
\begin{tabular}{lcr} 
\includegraphics[width=0.3\textwidth,height=0.3\textwidth]{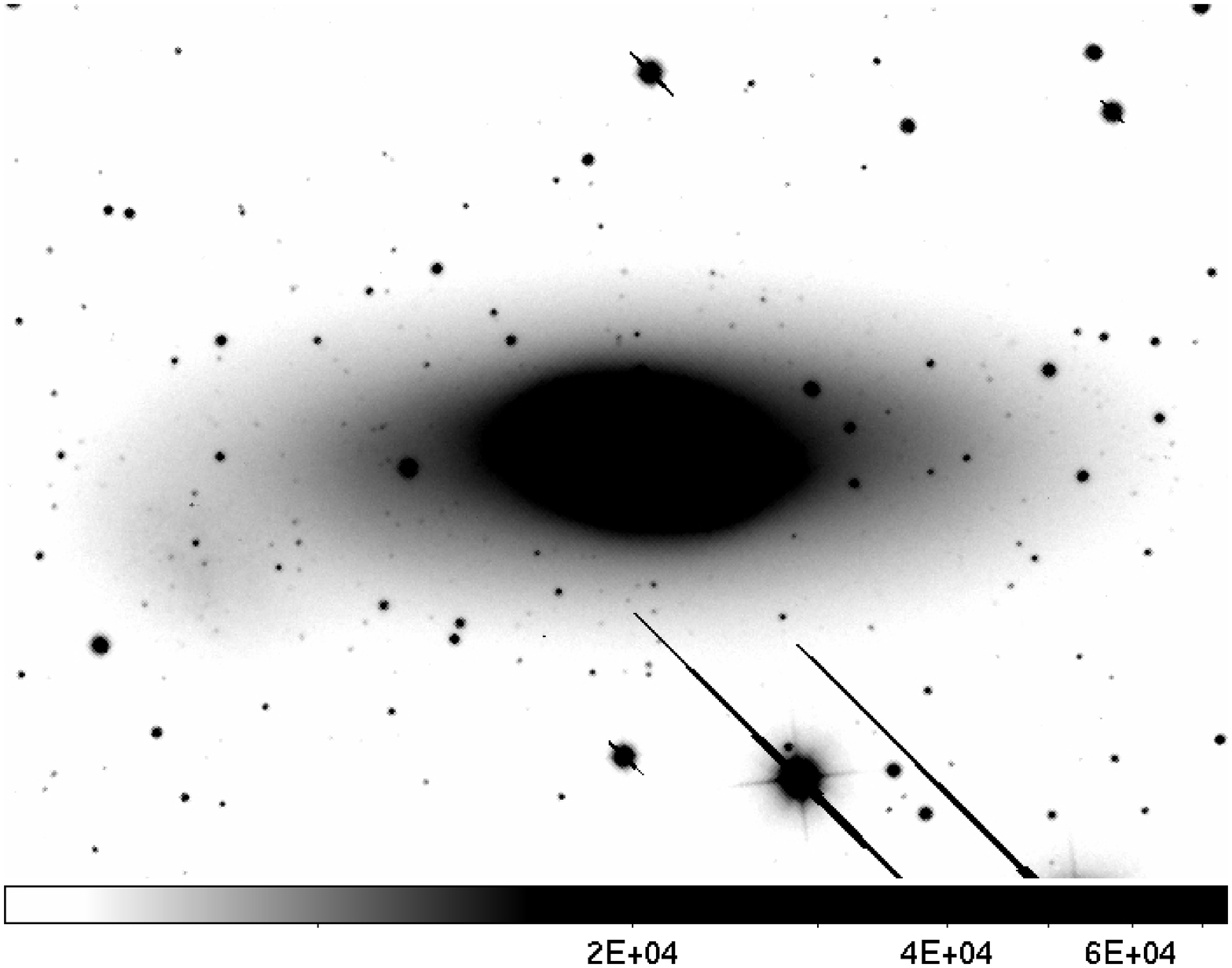}
\includegraphics[width=0.3\textwidth,height=0.3\textwidth]{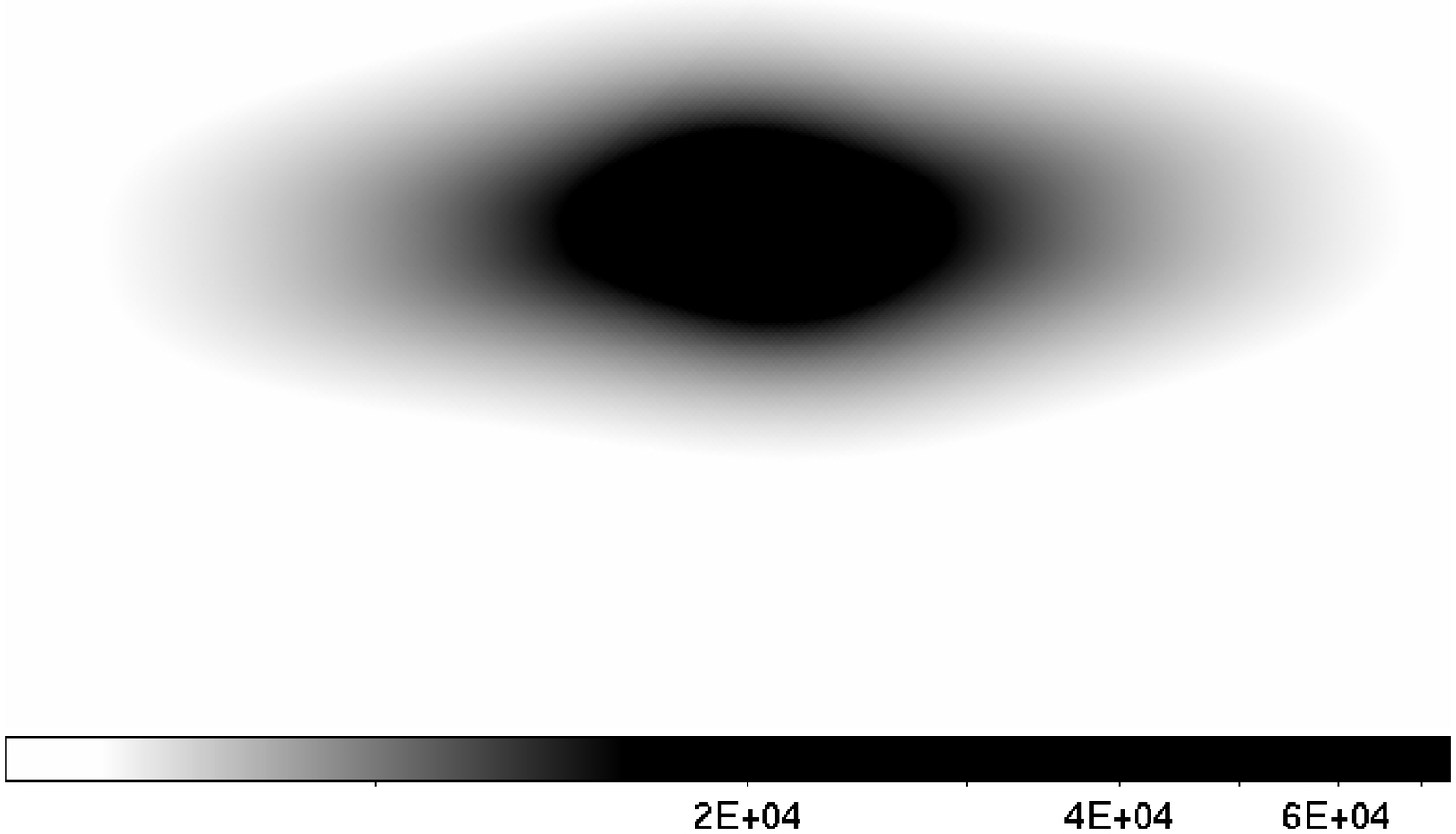}
\includegraphics[width=0.3\textwidth,height=0.3\textwidth]{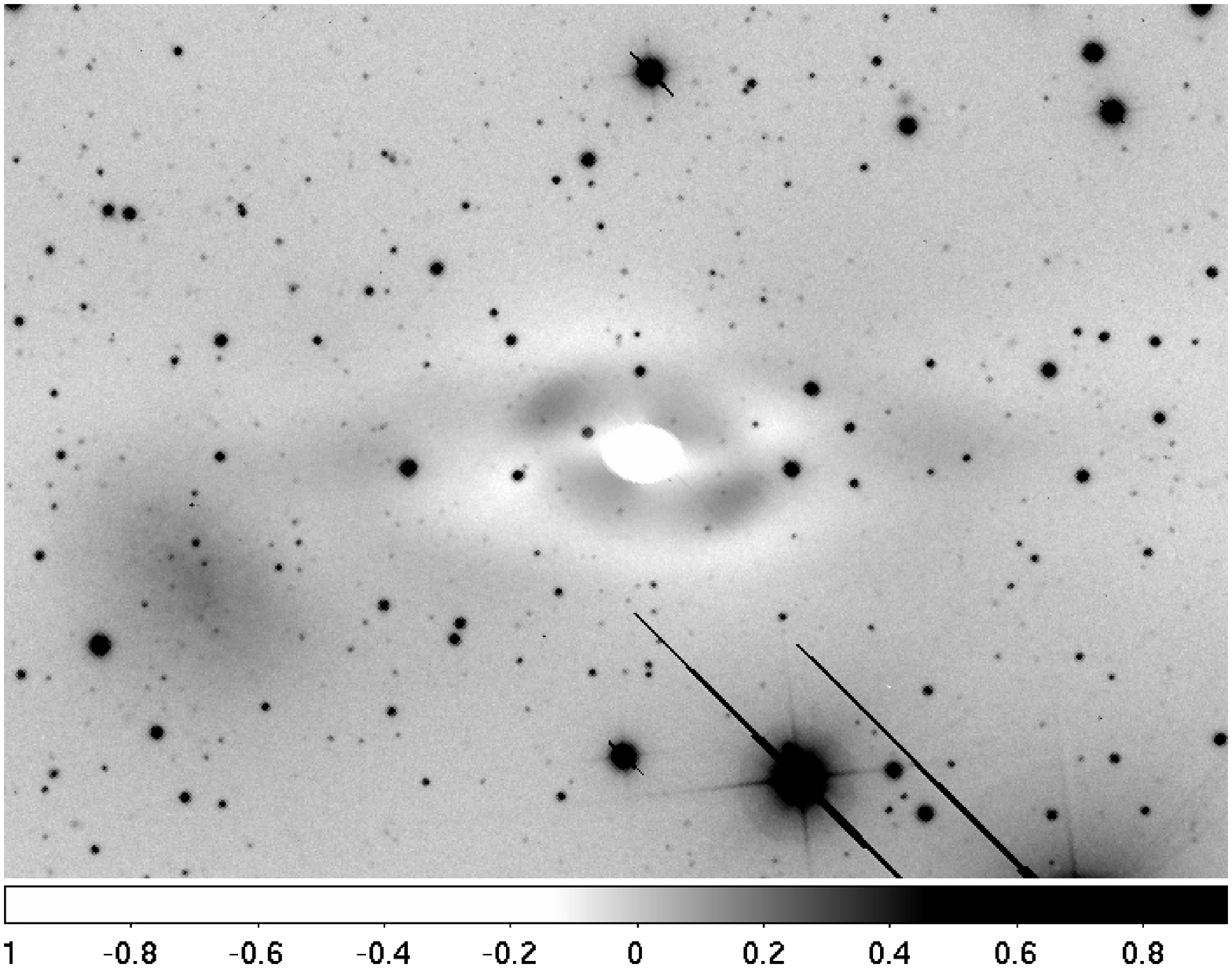}\\
\includegraphics[width=0.3\textwidth,height=0.3\textwidth]{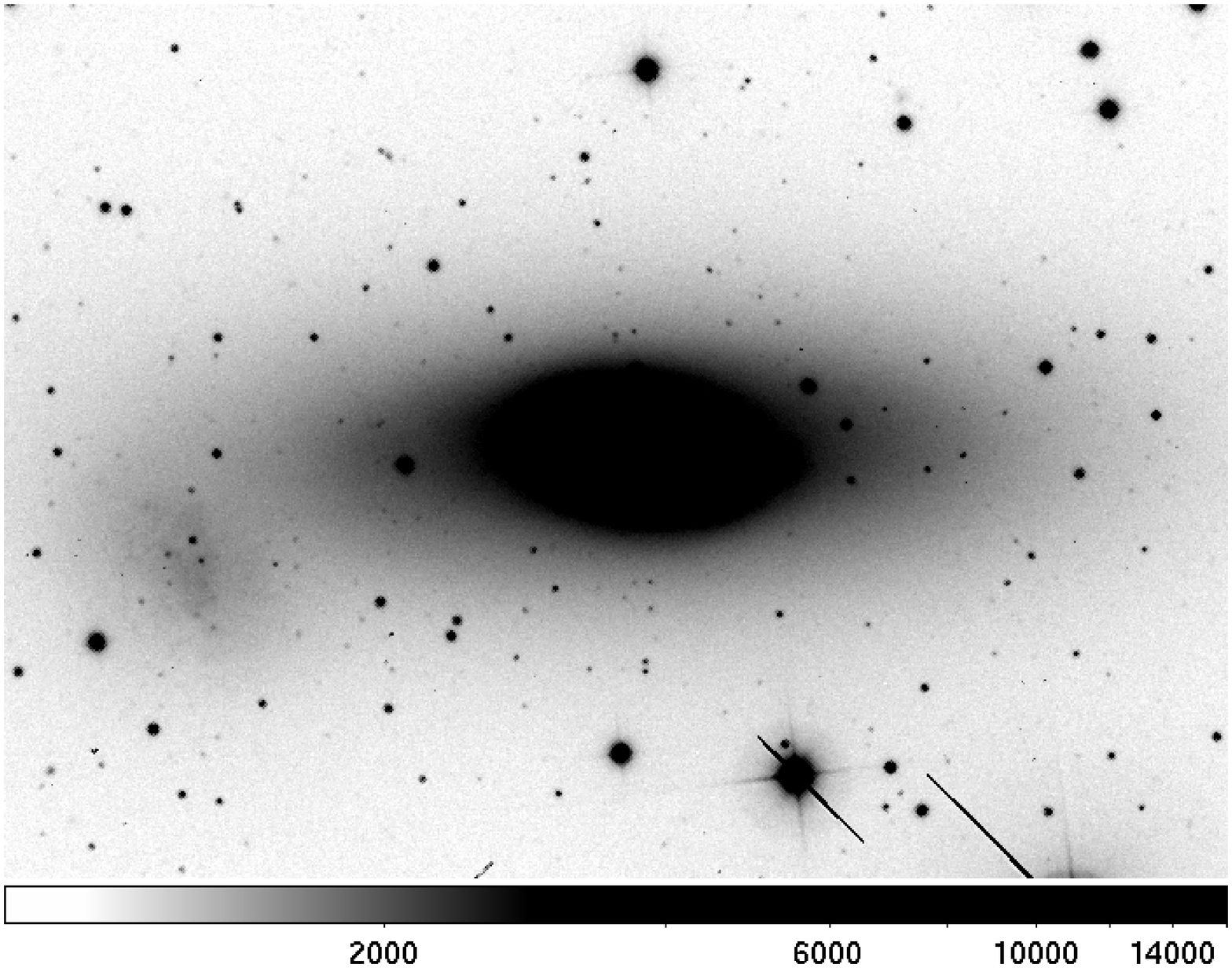}
\includegraphics[width=0.3\textwidth,height=0.3\textwidth]{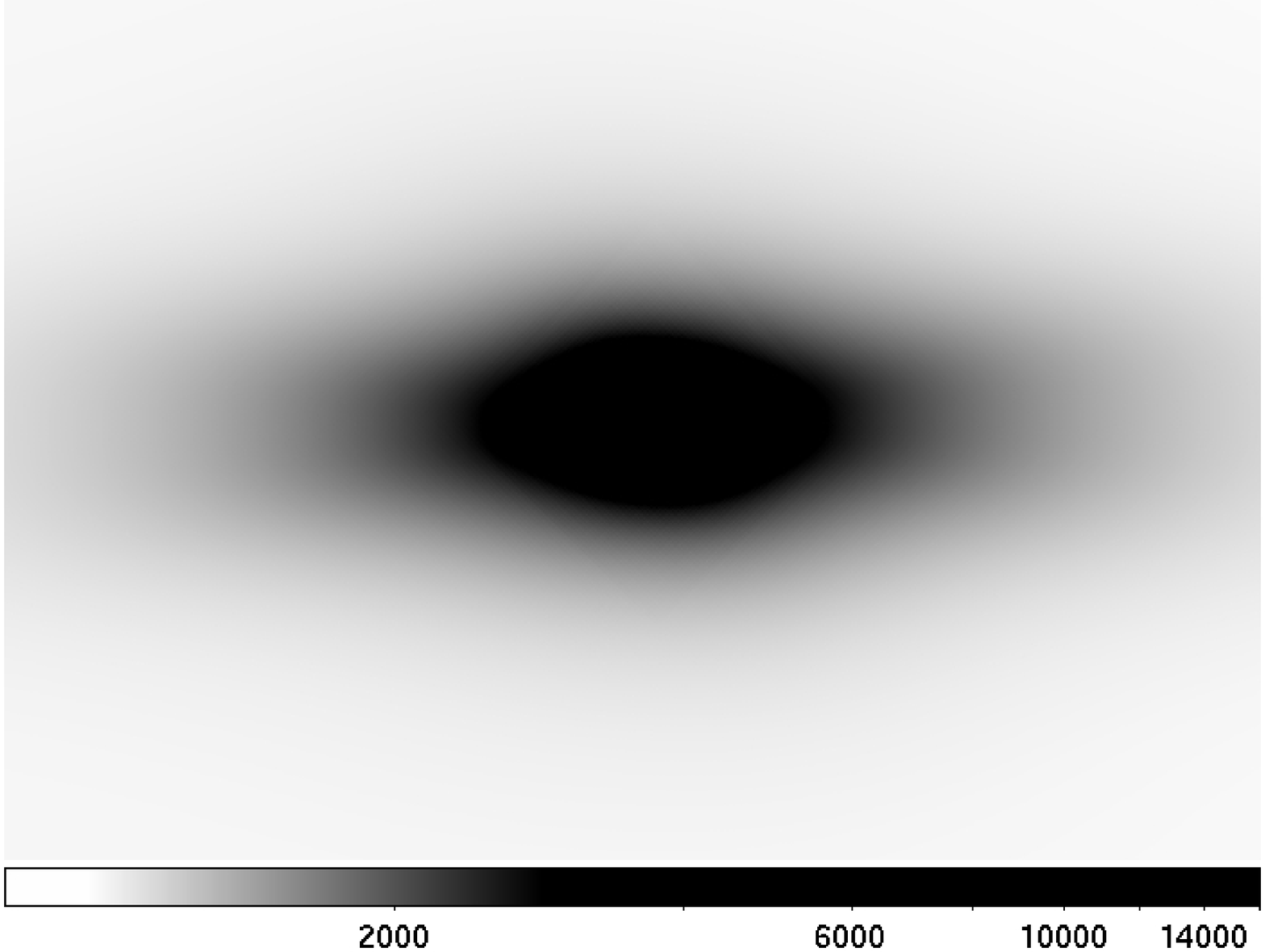}
\includegraphics[width=0.3\textwidth,height=0.3\textwidth]{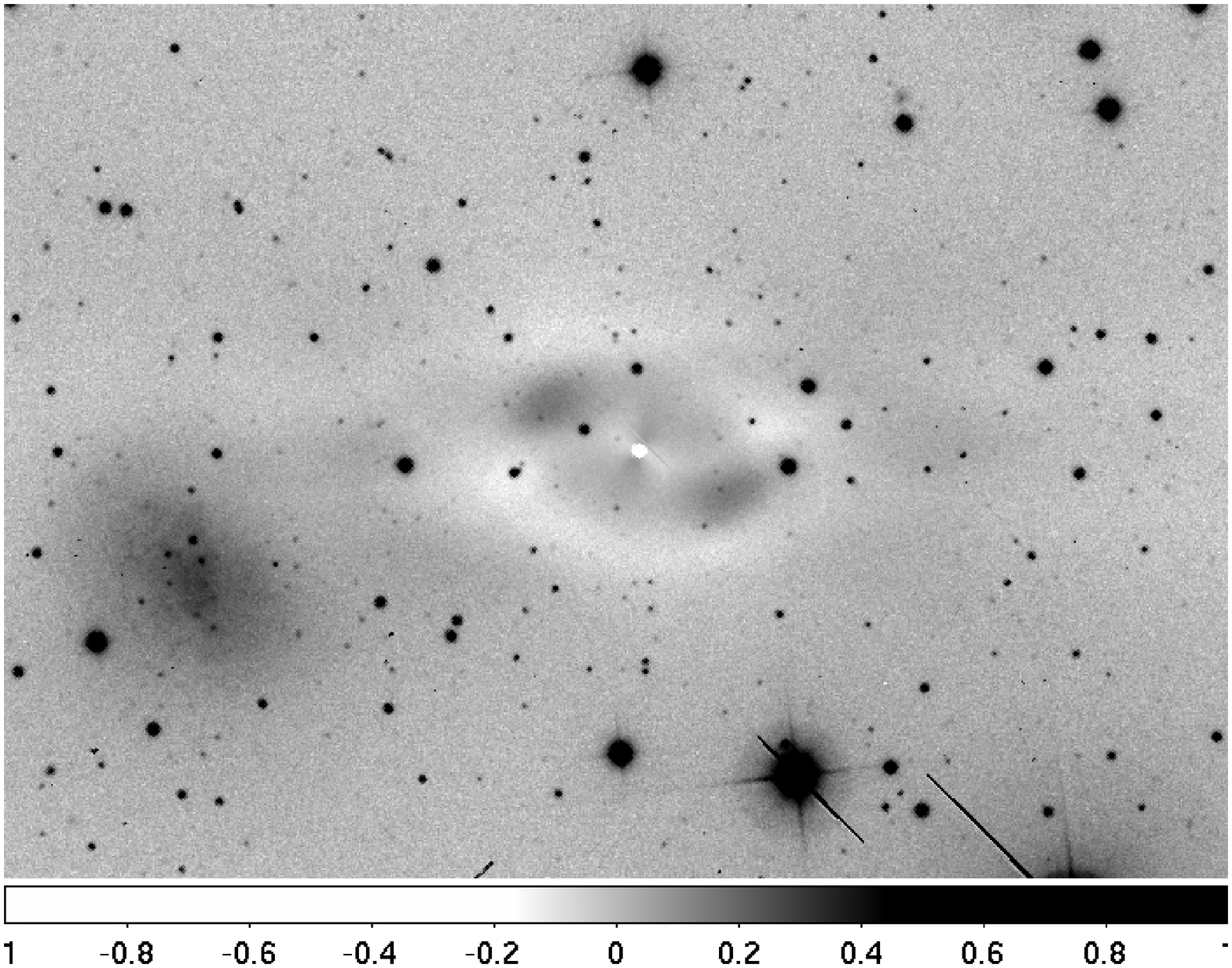}\\
\end{tabular}
\caption{GALFIT analysis of NGC~1023.  The top row is for a R-band
  image; the bottom row is for an B-band image. The data are obtained with the Isaac Newton Telescope on 1995 December 25, using the prime focus camera. The $300$ second exposures cover a field of view of $10.2' \times 10.2'$ \citep{EDO}. The first column shows the original data, the second column presents the model, and the third column shows the result of subtracting the model from the data. The residual image has been normalized by the original data, so that the values range from $0$ for a perfect fit, to $1$ or $-1$, where the data has not been fitted at all. \label{fig:companion}}
\end{figure*}

The first step toward modeling the spheroidal component is to
decompose continuum images into disk and spheroidal components, which
we have done using GALFIT \citep{Peng} to fit an exponential disk and
a de Vaucouleurs law spheroid to deep images of NGC~1023.
Figure~\ref{fig:companion} shows the result of this process carried
out on both B and R band images.  The companion galaxy NGC~1023A is
mildly apparent even in the raw images, but shows up very clearly in the
residuals when the model is subtracted.  The residual image also shows
evidence for the bar at the centre of this galaxy noted by
\citet{Deb}; since this faint feature is relatively localized at small
radii, and we are primarily interested in the balance between disk and
spheroid light at large radii, we do not attempt to model it any
further.  We could also have considered other models for the spheroid,
such as incorporating distinct bulge and halo components, or picking a
more general Sersic profile for the light distribution.  However, as
we have seen in Section~\ref{sec: Application to a model galaxy}, the
results are relatively insensitive to such subtleties.  Since the
primary goal is just to model to a reasonable approximation the
fraction of the light in the different components at different
positions, there is little benefit in trying to distinguish between
what are likely to be near-degenerate alternative fits.

\begin{figure}
\includegraphics[width=0.35\textwidth, angle=270]{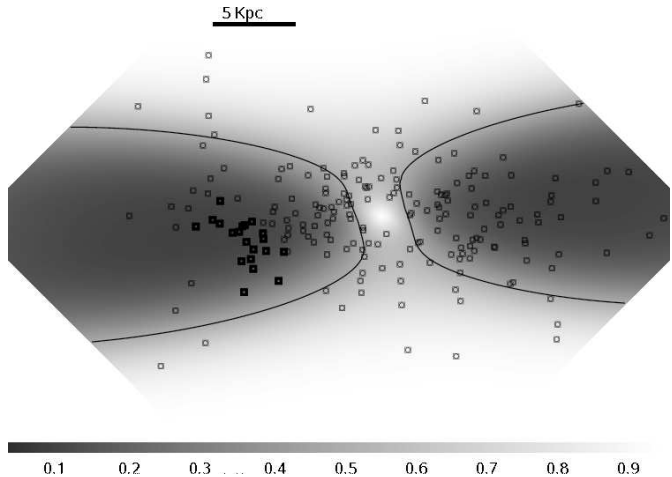}
\caption{Map showing the probability, $f$, that a PN at any given
  location is drawn from the spheroidal component of NGC~1023.  The
  positions of the detected PNe are superimposed. Bold squares identify NGC1023A PNe. The $f=0.5$ contour is shown in black. \label{fig:fwithPNe}}
\end{figure}
From the models in the two bands, we can calculate colors for the
individual components, and in this case we find $B - R = 1.62$ for the
disk and $1.61$ for the spheroid.  Thus there appears to be
essentially no color difference between the components.  We also find
no significant difference between the scale-lengths of the components
in the two bands: the best-fit model disk has a scale length of 67
arcseconds for the B-band data and 66 arcseconds for the R-band;
similarly, the spheroid has an effective radius of 24 arcseconds in
the B-band and 21 arcseconds in the R-band.  There is thus no evidence
for color gradients within each component.  This lack of color
variation between and within components makes the translation from
stellar continuum properties to the probability that a PN belongs to
the spheroid or disk particularly simple in this case, as the
probability is just the ratio of the component to total light at each
point. The bulge to total light ratio is 0.31 in the R-band and 0.35
in the B-band. These values compare well with estimates from the
cruder one-dimensional bulge--disk light decomposition performed in
\citep{EDO}, where the bulge-to-total light ratio was found to be
0.36. The two-dimensional map of the fraction of light in the
spheroidal component, $f$, as a function of position on the sky is
presented in Figure~\ref{fig:fwithPNe}.  As discussed in
Section~\ref{sec:decompose}, the value of $f$ clearly varies strongly
with azimuth, with the spheroid being the dominant component at all
radii close to the minor axis, underlining the necessity of this full
two-dimensional decomposition.

\subsection{The disk $+$ spheroid model}

\begin{figure}
\includegraphics[width=0.45\textwidth]{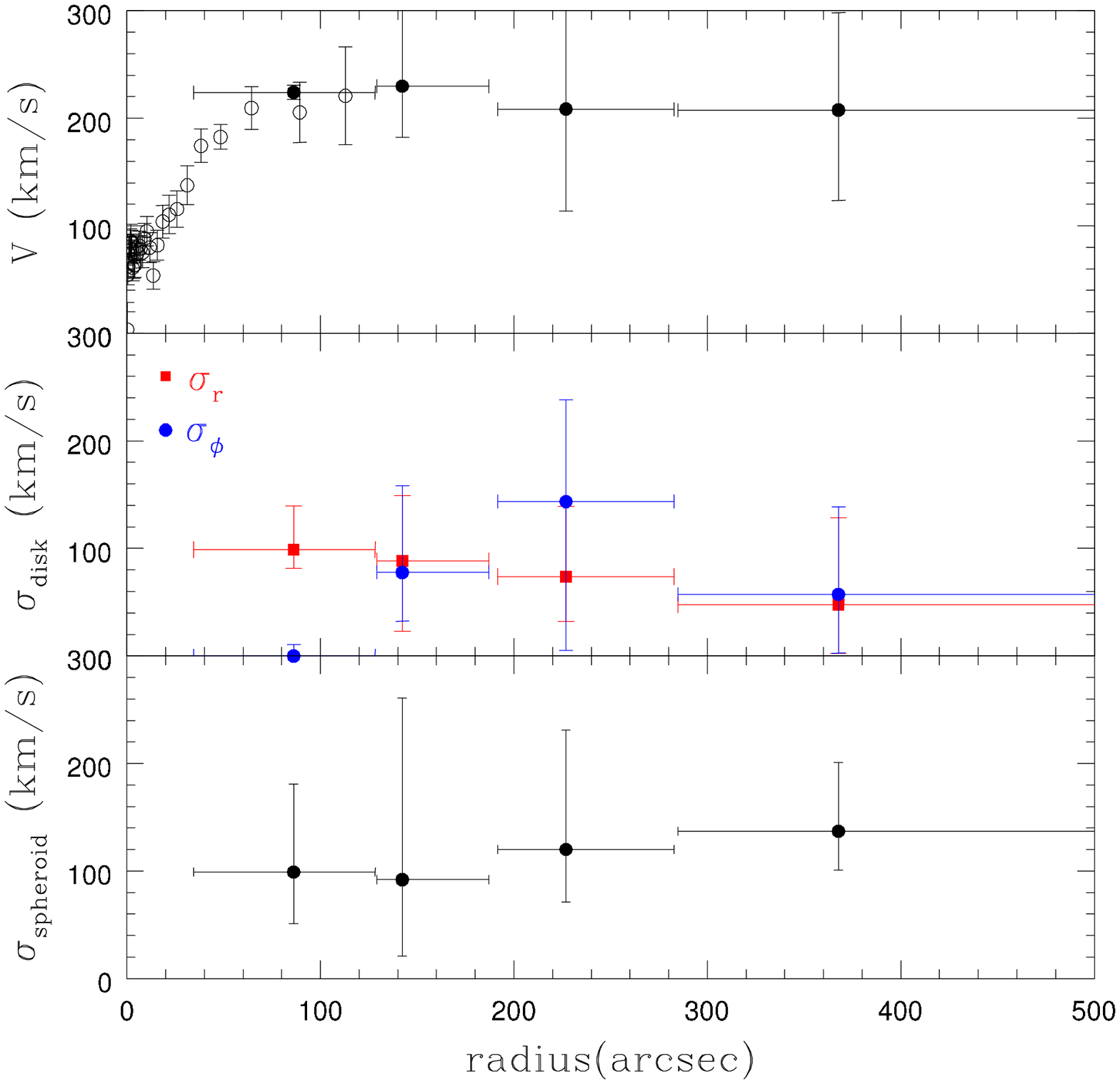}
\caption{Derived mean rotation speed, the components of the disk 
  velocity dispersion, and the spheroid dispersion versus radius for
  NGC~1023. The filled symbols are from the maximum-likelihood
  analysis, with vertical error bars indicating uncertainty and
  horizontal error bars showing the extent of each radial bin (with
  the point plotted at the median location for a PN in that bin.  The
  open symbols show rotation velocities derived from absorption-line
  data by \citet{Deb}.\label{fig:like-h}}
\end{figure}

Having calculated the division between spheroid and disk light, we can
now carry out the likelihood analysis for NGC~1023 incorporating this
extra component, as set out in Section~\ref{sec:diskandspheroid}. the
resulting best-fit kinematic parameters as a function of radius are
shown in Figure~\ref{fig:like-h}.  We now start to see the
characteristic properties of a normal disk galaxy.  Rotation dominates
random motions in the disk at all radii, and the mean rotation speed
stays approximately constant out to the last points shown.  In this
plot, we have also filled in the missing stellar rotation velocity at
small radii from conventional absorption-line data, and it is clear
that the two techniques agree extremely well where they overlap.  The
spheroid dispersion profile is also very well-behaved, showing little
if any variation with radius (justifying the compromise in radial
binning discussed in Section~\ref{sec:diskandspheroid}).  It is also
notable that the approximately-constant spheroid dispersion is
consistent with a value of $V/\sqrt{2}$, which is what one would
obtain for the simplest possible model of a singular isothermal sphere
potential, in which circular speed and velocity dispersion are related
in this way.  This consistency is reassuring, as the fitting procedure
in no way imposed it on the results.

As one further enhancement, one might also expect some rotational
velocity, $V_s$, in the spheroidal component, which might affect the
results.  Indeed, there is known to be a strong correlation between
ellipticity and rotational speed in low-luminosity spheroidal systems
like this bulge  \citep[see Figure 4.14 in][]{Binney}.  The GALFIT modeling shows that the
spheroid in NGC~1023 has an ellipticity of $\sim 0.25$, which
translates into a predicted value of $V_s/\sigma_s \sim 0.5$.  We have
experimented with including rotation at this level in the spheroidal
component by modifying the first term in
Equation~\ref{eq:dplussmodel}, but find that at this level the
inclusion of rotation makes no significant difference to the remaining
kinematic parameters.

\begin{figure}
\includegraphics[width=0.45\textwidth]{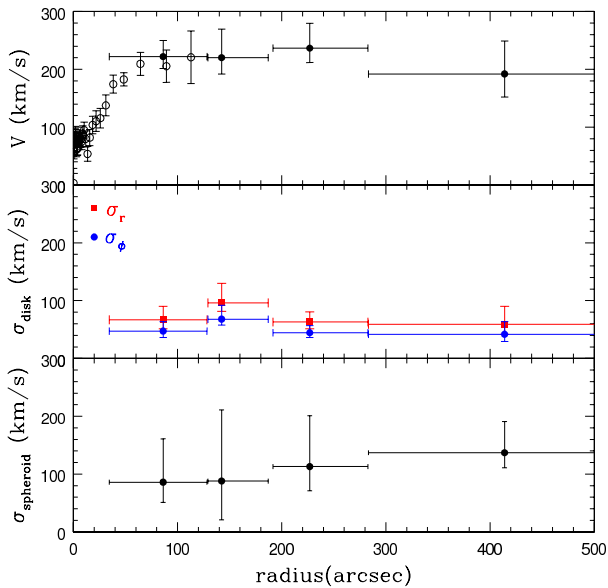}
\caption{As for Figure~\ref{fig:like-h}, but with the ratio of disk
  dispersions components constrained by the epicyclic
  approximation.\label{fig:like-h-epi}}
\end{figure}
The only slight disappointment in the fitting process is that the
ratio between the components of disk dispersion is somewhat less well
defined than was the case in the disk-only model fit, which presumably
reflects the impact of the extra spheroidal dispersion free parameter in
this model.  However, now that we clearly have a well-behaved normal
disk system in this galaxy, it seems reasonable to reduce the number
of free parameters by invoking the epicyclic approximation.  In
particular, for such a cold disk system with a flat rotation curve, we
expect $\sigma_\phi/\sigma_r = 1/\sqrt{2}$.  With this additional
constraint, we obtain the kinematic parameters plotted in
Figure~\ref{fig:like-h-epi}.  The error bars on all parameters are
duly reduced, and we now find that the kinematics of NGC~1023 are
exactly what one would expect for a very normal disk galaxy, with
$V / \sigma_{\phi} \simeq 4.1$ throughout the disk, similar to
what one finds in the stellar component of a spiral galaxy.  Thus, we
seem to have found the explanation for the strange kinematics
originally inferred from these data for NGC~1023, and it is now
revealed to be most likely a spiral galaxy that has simply been stripped
of its gas.

\subsection{PNe objects rejections: the footprint of an ongoing merger }
\label{sec: Rejections}

\begin{figure}
\includegraphics[width=0.45\textwidth]{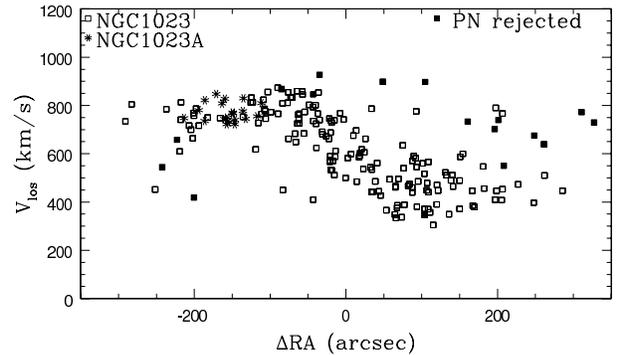}
\caption{Plot of radial velocity versus right ascension for the PNe in
  NGC~1023.  Filled symbols show the objects rejected by the
  likelihood clipping in the full disk+spheroid model.  Asterisks show
  the PNe attributed to NGC~1023A.\label{fig:rejects}}
\end{figure}
However, the story does not quite finish there.  Even with the full
disk+spheroid kinematic model, 17 PNe are still likelihood-clipped at
a threshold probability of $2.1 \sigma$.  In Figure~\ref{fig:rejects},
we show the locations of the PNe in both velocity and right ascension,
with the rejected objects highlighted.  Since NGC~1023 lies at a
position angle very close to 90 degrees, the spatial coordinate is
essentially the distance along the major axis, so this is a
conventional position--velocity diagram, with the usual antisymmetric
signature of a rotating disk evident in the PNe.  The rejected PNe,
however, do not show such antisymmetry, with the vast majority located
on one side of the galaxy.  Such an asymmetric arrangement is clearly
not consistent with errors arising from a poor choice for any
axisymmetric element in the model, or from mis-identified unrelated
background objects.

A clue to their origin comes from considering the location of
NGC~1023A in this plot, as the rejected PNe seem almost all to form a
continuous stream that passes through this companion galaxy, as one
might expect if the systems are tidally interacting
\citet{Capaccioli}.  It therefore seems likely that these PNe lie in
the tidal debris from this companion as it is stripped in an on-going
minor merger.

\section{Conclusions}
\label{sec: Conclusions}

We have presented a new method for analyzing the kinematics of disk
galaxies as derived from individual stellar tracers such as PNe.  This
hybrid technique bins data radially in the galaxy to maintain the
maximum flexibility in the inferred kinematics, but uses a likelihood
analysis within each bin so as to derive the maximum amount of
information from the discrete data points.   In addition,
we use photometric decomposition of continuum images to assign a
probability to each kinematic tracer of belonging to either the disk
or the spheroidal component of the galaxy being modeled.

Application of the method to simulated data shows that it reproduces
most of the intrinsic dynamics of a galaxy even when the number of
discrete kinematic data points is relatively modest.  Application of
this technique to NGC~1023 has offered an explanation for the strange
kinematics previously inferred for this system.  These peculiar
properties can be entirely attributed to uncorrected contamination by
the galaxy's spheroidal component, and when this element is properly
modeled, the galaxy is revealed to have stellar kinematics entirely
consistent with a normal spiral system, with a cold
rotationally-supported disk and a hot spheroid with the expected
velocity dispersion.  This result favours a model in which NGC~1023
formed from a spiral via a simple gas stripping process or secular
evolution, rather than through a more disruptive minor merger.

One benefit of the likelihood analysis made clear by this application is
that it is possible to ``likelihood clip'', to identify objects that do
not seem to be drawn from the model, in a reasonably robust and
objective manner.  Comparison of Figures~\ref{fig:like-h-epi} and
\ref{fig:like-nh-clip} shows that such clipping can deal quite well
with even the rather extreme case where only the disk is included in
the model.  Where a more realistic disk+spheroid model is adopted, it
seems to do a good job of identifying features like stellar streams,
adding to our understanding of the dynamical properties of such
systems.

Clearly, though, although the detailed dynamics of this galaxy are
interesting, one cannot infer any general conclusions about the
formation of S0 galaxies from a single object.  Therefore, the next
step in this project will involve applying this analysis technique to
observations of a larger sample of S0 galaxies whose PNe kinematics
have been observed with PN.S. By obtaining a measure of the stellar
kinematics of S0s in regions spanning a range of galaxy densities, we
will be able to see if they all have the stripped-spiral properties of
NGC~1023, and hence whether there is a single route to S0 formation
irrespective of environment.

\section*{Acknowledgements}
We would like to thank Boris Haeussler for his extensive support in
the use of GALFIT, and the anonymous referee for significantly
improving the presentation of this paper.  K. Saha also thanks the
Alexander von Humboldt foundation for its financial support.

\end{document}